\documentclass[aps,prd,preprint,showpacs,eqsecnum,superscriptaddress,nofootinbib]{revtex4}

\usepackage{graphicx,amssymb}

\def\fig#1{fig.~{\ref{#1}}}
\def\figs#1#2{figs.~{\ref{#1}} and {\ref{#2}}}

\begin{document}

\noindent
\hfill Brown-HET-1551

\vskip 1 cm

\title{Leading Singularities of the
Two-Loop Six-Particle MHV Amplitude}

\author{Freddy Cachazo}

\affiliation{Perimeter Institute for Theoretical Physics,
Waterloo, Ontario N2J 2W9, Canada}

\author{Marcus Spradlin}

\affiliation{Brown University, Providence, Rhode Island 02912, USA}

\author{Anastasia Volovich}

\affiliation{Brown University, Providence, Rhode Island 02912, USA}

\begin{abstract}

We use the leading singularity technique to determine the planar
six-particle two-loop MHV amplitude in ${\cal N}=4$ super Yang-Mills
in terms of a simple basis of integrals. Our result for the parity
even part of the amplitude agrees with the one recently presented
in~\cite{Bern:2008ap}.
The parity-odd part of the amplitude is a new result.
The leading singularity
technique reduces the determination of the amplitude to finding the
solution to a system of linear equations. The system of equations is
easily found by computing residues. We present the complete system
of equations which determines the whole amplitude, and solve the
two-by-two blocks analytically.  Larger blocks are solved
numerically in order to test the ABDK/BDS
iterative structure.

\end{abstract}


\maketitle

\section{Introduction}

Scattering amplitudes in massless gauge theories are remarkable
objects with many properties hidden in the complexity of their
Feynman diagram expansion.
It has been known for decades that much information about an
amplitude can be gleaned just from knowing the structure of its
singularities (see~\cite{SMatrix}).
In Yang-Mills theories, extensive use of branch cut singularities
was shown to tame much complexity in the computation
of loop amplitudes via techniques that came
to be known as the unitarity based
method~\cite{Bern:1994zx,Bern:1994cg,Bern:1995db,Bern:1996je,BDDKSelfDual,Bern:1997sc,Bern:2004cz}.

One of the most surprising features of Yang-Mills theory and gravity
is that their tree level amplitudes can be completely determined by
exploiting only their behavior near a subset of their
singularities~\cite{BCFTree,BCFW,ArkaniHamed:2008yf}.
Another surprise, this time in
${\cal N}=4$ SYM, is that the problem of computing one-loop
amplitudes can be reduced to that of computing tree
amplitudes~\cite{BCFLoop}. The key to such striking simplification
is that although the loop amplitude possesses poles and many branch
cuts with a complicated structure of intersections, it is completely
determined by the structure of the highest codimension
singularities. These are known as the leading
singularities~\cite{SMatrix} and are computed by cutting all
propagators in the diagram.

Applying the same technique at higher loops was first attempted
in~\cite{Buchbinder:2005wp} and refined
in~\cite{Bern:2007ct,Cachazo:2008dx}. In~\cite{Buchbinder:2005wp}
and in~\cite{Bern:2007ct} both the leading singularity as well as
subleading ({\it i.e.}, lower codimension) singularities were used to
constrain the form of higher loop amplitudes.
In~\cite{Cachazo:2008vp}, the leading singularity was
shown to be much more powerful than expected. It turns out that in
massless theories, whenever all propagators are cut, one is actually
studying many isolated singularities at the same time. The proposal
of~\cite{Cachazo:2008vp}, building on observations made
in~\cite{Cachazo:2008dx}, is to use each isolated singularity
independently.

The new leading singularity technique
outlined in~\cite{Cachazo:2008vp} has three notable features.
Firstly,
for any amplitude under consideration it builds a natural set of integrals,
which we call
the geometric integrals, that can be used to construct a basis.
Secondly, the coefficients of the integrals
can be determined by solving {\it linear} equations.
Finally, these linear equations are easily obtained by
computing residues using Cauchy's theorem.
The utility of this new
technique was demonstrated in~\cite{Cachazo:2008vp},
where it was shown that it easily reproduces the
result for the two-loop
five-particle amplitude in ${\cal N}=4$ SYM previously
computed in~\cite{TwoLoopFiveB} using the unitarity based
method.

In this paper we apply the leading singularity technique to a
much more challenging case, the planar two-loop six-particle
MHV amplitude ${A}^{(2)}_{6,{\rm MHV}}$.
The parity-even part of the normalized amplitude
${A}^{(2)}_{6,{\rm MHV}}/{A}^{\rm tree}_{6,{\rm MHV}}$
was computed recently in~\cite{Bern:2008ap} using the
unitarity based method. This was already an impressive display
of computational power.  In the five particle case studied
in~\cite{TwoLoopFiveA,TwoLoopFiveB}
the parity-odd part is noticeably of a higher
degree of complexity than the parity-even part, and there is no reason
to suspect that this would not be the case also for $n=6$.
Here we find that the full coefficients (both the even and odd parts) emerge
by solving the relatively simple linear equations presented
explicitly below.

For six particles we find a new phenomenon which was not encountered
in the $n=4,5$ cases studied in~\cite{Cachazo:2008vp}:
while there are 177 geometric integrals, the equations only
fix 159 linear combinations of their coefficients, leaving 18 linear
combinations undetermined.
The reason is that the set of geometric integrals is overcomplete, so we cannot
have expected to find a unique solution.
One can show that by using well-known
techniques~\cite{Melrose:1965kb,vanNeerven:1983vr}
it is possible to build linear
relations between seemingly independent integrals. In
section IV we analyze the  relevant reduction identities and
identify 18 relations amongst the integrals
in the geometric basis.  Thus there is no ambiguity beyond that required
by reduction identities, so we conclude that ${ A}_{6,{\rm MHV}}^{(2)}$
is in fact completely
determined by its leading singularities.

It is important to stress that the leading singularity method turns
loop integrals into contour integrals which are finite in four
dimensions and knows nothing about how one might choose to regulate
the infrared divergences that typically appear when carrying out the loop
integrals. In dimensional regularization, amplitudes occasionally
contain additional terms in the integrand which vanish in $D=4$.
These so-called ``$\mu$-terms'' (see for example~\cite{Bern:2002tk}
for a thorough treatment)
cannot be detected by the leading
singularity in $D=4$.

One motivation for computing the MHV six-particle amplitude,
beyond its serving as a testing ground for the
leading singularity method,
is to study the proposed
iterative relation between MHV loop amplitudes known as the
ABDK/BDS ansatz~\cite{ABDK,BDS}.
The ansatz has been shown to hold
for four particles up to three loops~\cite{ABDK,BDS} and
for five particles up to two loops~\cite{TwoLoopFiveA,TwoLoopFiveB}.
However, it was shown to
break down for the parity-even part of the two-loop six-particle MHV
amplitude in~\cite{Bern:2008ap}. In this paper we
find numerical evidence that the parity-odd part
of the amplitude does satisfy the ABDK/BDS ansatz.

\section{Outline of the Calculation}

The object of interest is the planar six-particle two-loop
MHV amplitude in ${\cal N} = 4$ super Yang-Mills.
The goal is to find a compact
expression for this amplitude as a linear combination of relatively simple
integrals.  The leading singularity method~\cite{Cachazo:2008vp}
provides both a natural set of integrals to work with, as well as
a system of linear equations which determine the coefficients of
those integrals.

In this section we provide a detailed outline of the steps involved
in setting up the calculation.
The first three subsections are relevant to NMHV as well as MHV
amplitudes, since the homogeneous part of the system of linear
equations is helicity independent.
In subsection II.D we compute the inhomogeneous terms for the MHV helicity
configuration.
The final linear equations which determine
the coefficients of the MHV amplitude are presented explicitly
in section III.

\subsection{Review of the Leading Singularity Method}

Suppose we are interested in calculating some $L$-loop scattering amplitude
${ A}$.
On the one hand, the amplitude may of course be represented
as a sum over Feynman diagrams $F_j$,
\begin{equation}
\label{eq:one}
{ A}(k) = \sum_j \int \prod_{a=1}^L d^d \ell_a\
F_j(k,\ell)\,,
\end{equation}
where $k$ are external momenta and $\ell_a$ are the loop momenta.
However it is frequently the case, especially in theories
as rich as ${\cal N} = 4$ SYM,
that directly calculating the sum over Feynman diagrams would be impractical.
Rather the calculation proceeds
by expressing ${ A}$ as a linear combination of relatively simple integrals
in some appropriate basis $\{I_i\}$,
\begin{equation}
\label{eq:two}
{ A}(k) = \sum_i c_i(k) \int \prod_{a=1}^L d^d \ell_a\
I_i(k,\ell)\,,
\end{equation}
and then determining the coefficients $c_i$ by other means, such as
the unitarity based method.
If the set of integrals $\{I_i\}$ is overcomplete, then the coefficients
$c_i(k)$ are not uniquely defined.

The basic idea underlying the leading singularity method is that
the sum over Feynman diagrams in~(\ref{eq:one}) possesses
singularities which must be properly reproduced by any
representation~(\ref{eq:two})
of the amplitude
in terms of simpler integrals.
At the same time, any singularities in the set of integrals
which are not present in the sum over Feynman diagrams must be spurious.

The most common kind of singularities in Feynman diagrams are poles,
associated to collinear or multi-particle singularities, and branch
cuts, associated to unitarity cuts. These branch cuts can themselves
possess branch cuts leading to higher codimension singularities. The
latter are computed by cutting propagators or
equivalently~\cite{Cachazo:2008dx} by promoting the loop integral to
be a contour integral. (The observation that the Lorentz
invariant phase space integral of a null vector can be recast
as a contour integral was first discussed in~\cite{CSW}.) The
contour is chosen to reproduce the behavior of the delta-functions
in the cut calculations. In general, this gives rise to contours
which compute the residue on several isolated singularities at the
same time.

For example, consider the one-loop massless scalar box.
Replacing each propagator by a delta-function leads us to consider
the integral
\begin{equation}
\int d^4 \ell\,
\delta(\ell^2)
\delta((\ell - k_1)^2)
\delta((\ell - k_1 - k_2)^2)
\delta((\ell + k_4)^2)\,.
\end{equation}
For generic external momenta $k_i$ these delta-functions localize
the $\ell$ integral onto two discrete points in complex
$\ell$-space ($\mathbb{C}^4$).
With the leading singularity method we do
not
replace propagators by delta-functions,
but rather we consider two separate
$T^4$ contours in $\mathbb{C}^4$,
each of which computes the
residue of the integrand
on only one of the two isolated singularities.
Then by equating~(\ref{eq:one}) and~(\ref{eq:two})
and performing the integral
\begin{equation}
\label{eq:method}
\sum_i c_i(k) \int_\Gamma d^4 \ell\, I_i(k, \ell)
=
\int_\Gamma d^4 \ell\,
 \sum_j F_j(k, \ell)
\end{equation}
we obtain one linear equation on the coefficients $c_i$ for
each contour $\Gamma$.
At $L$ loops each contour is a $T^{4L}$ inside $\mathbb{C}^{4 L}$.
Since the number of isolated singularities in a generic
$L$-loop diagram can be as high as $2^L$ (simple diagrams can have
fewer isolated singularities), the leading singularity method gives
rise to an exponentially large (in $L$) number of linear equations
for the coefficients $c_i$.
We note that the homogeneous part of these linear equations (the left-hand
side of~(\ref{eq:method})) depends only on the set of integrals $\{I_i\}$ and
the choice of contours, while
the details of which particular helicity configuration
is under consideration enters
only into the inhomogeneous terms on the right-hand side.

\subsection{Choosing Useful Contours}

The formula~(\ref{eq:method}) gives a linear equation
on the coefficients $c_i$ for any contour $\Gamma$ in ${\mathbb{C}}^{4 L}$.
Of course if we choose some random contour $\Gamma$ then we will
typically get the useless equation $0=0$.  In order to get useful
equations we should use $T^{4 L}$ contours which calculate residues at the
known singularities of the right-hand side.
It is clear that
in the sum over Feynman
diagrams, singularities occur when internal propagators go on-shell.

\begin{figure}
\includegraphics{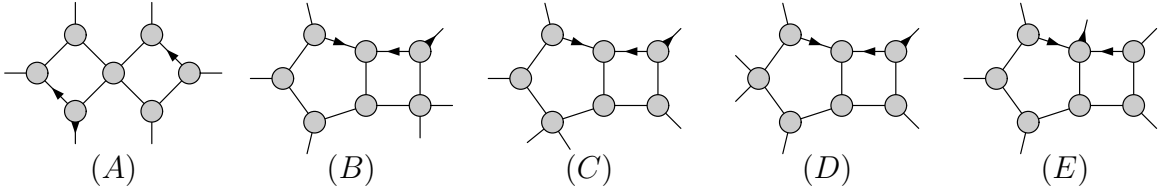}
\caption{The five indepedent 8-propagator topologies.
Each diagram represents a sum of those Feynman diagrams
in which all of the 8 indicated propagators are present.
In each diagram the external momenta are labeled clockwise
beginning with $k_1$ at the position of the arrow.
Also in each diagram $p$ is the loop momentum  in the left loop and $q$ is the
momentum in the right loop.
}
\label{EightPropagatorTopologies}
\end{figure}

For six particles at two-loops there are two classes of useful $T^8$ contours.
The most obvious $T^8$ contours are those which are chosen to calculate
the residue at points in $\mathbb{C}^8$
where eight propagators go on-shell simultaneously.
These contours are associated with the five different topologies
shown in~\fig{EightPropagatorTopologies}.  Actually each topology
in~\fig{EightPropagatorTopologies} is
a diagrammatic shorthand for four distinct $T^8$ contours.
For example, the singularities of Feynman diagrams
with topology $(D)$ are situated at
the locus
\begin{eqnarray}
S_{(D)} = \{ (p,q) \in \mathbb{C}^4 \times \mathbb{C}^4 &:&
p^2 = 0, ~ (p + k_6)^2 = 0, ~ (p + k_{456})^2 = 0, ~ (p - k_{12})^2 = 0,\cr
&&
q^2 = 0, ~ (q + k_1)^2 = 0, ~ (q + k_{12})^2 = 0, ~ (p + q)^2 = 0 \}\,.
\end{eqnarray}
For generic external momenta $k_i$ $S_{(D)}$ consists of four
distinct points in $\mathbb{C}^8$ of the form
$(p^{(i)}, q^{(j)})$ for $i,j=1,2$.
Correspondingly there are four different contours $\Gamma$ associated
with topology $(D)$, one which computes the residue
of the integrand at each of these
four isolated singularities.

\begin{figure}
\includegraphics{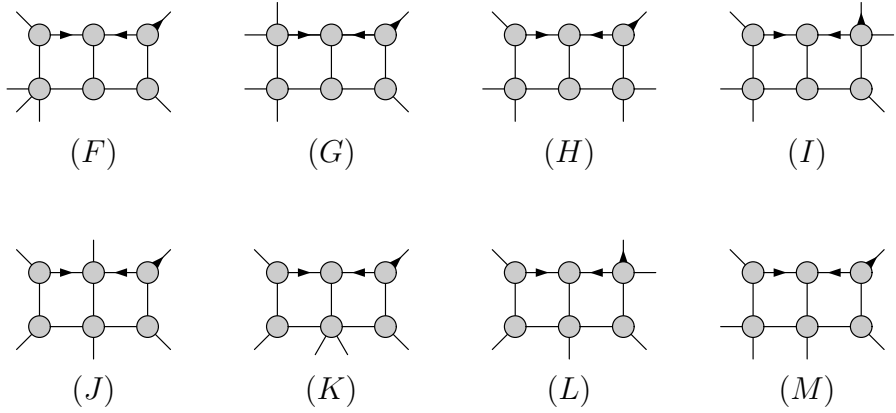}
\caption{The eight independent 7-propagator topologies.
See~\fig{EightPropagatorTopologies} for details.
}
\label{SevenPropagatorTopologies}
\end{figure}

The less obvious $T^8$ contours are those in which only seven propagators
are apparent but an eighth singularity appears due to a Jacobian.
These contours are associated with the eight different topologies
shown in~\fig{SevenPropagatorTopologies}.
For example, let us consider topology $(F)$.  For fixed
loop momentum
$p$ the singularities
in the $q$ integral occur at the locus
\begin{equation}
S_{(F)q} = \{ q \in \mathbb{C}^4 : q^2 = 0, ~ (q + k_1)^2 = 0, ~
(q + k_{12})^2 = 0, ~ (q + p)^2 = 0 \}\,,
\end{equation}
which consists of two points $\{q^{(1)}, q^{(2)}\}$ in $\mathbb{C}^4$.
For each of these two singularities there is a contour $\Gamma_q$ such that
integrating $q$ over $\Gamma_q$ computes the residue at that singularity.
Integrating over either contour produces the same
Jacobian factor
\begin{equation}
\label{eq:jacobian}
\int_{\Gamma_q} d^4 q\ \frac{1}{q^2 (q + k_1)^2 (q+ k_{12})^2 (q + p)^2}
= {1 \over 2} {1 \over (k_1 + k_2)^2 (p - k_1)^2}\,.
\end{equation}
The new singularity $1/(p - k_1)^2$ combines with the three
remaining singularities manifest in topology $(F)$ so that the
integral over $p$ can be localized by integrating over contours which
compute the residue at the points
\begin{equation}
S_{(F)p} = \{ p \in \mathbb{C}^4 : p^2 = 0, ~ (p + k_6)^2 = 0, ~
(p - k_{12})^2 = 0, ~ (p - k_1)^2 = 0\}\,.
\end{equation}

We proceed analagously for each of the eight topologies
shown in~\fig{SevenPropagatorTopologies}. In each case
we first integrate the right-hand loop momentum $q$ and then use the
additional singularity generated by the Jacobian to integrate the
left-hand loop momentum $p$, thereby completely localizing the integral
onto a set of discrete points.

For the MHV amplitude it turns out that the linear equations generated
by the 13 types of contours described
in~\figs{EightPropagatorTopologies}{SevenPropagatorTopologies}
are sufficient to uniquely determine all coefficients,
although there certainly are additional contours that could be used
to generate additional equations from~(\ref{eq:method}).
We have checked that a class of additional equations are indeed
satisfied by the MHV coefficients, thus providing a strong consistency
check on the coefficients obtained from solving the equations in section III.

\subsection{The Geometric Integrals}

Although equation~(\ref{eq:method})
can be used to generate linear equations for coefficients
in an arbitrary basis $\{I_i\}$, the leading singularity method
suggests a natural set of integrals in which
the left-hand side is easy to
compute and the individual integrals
have a geometric interpretation, in a sense we now explain.

\begin{figure}
\includegraphics{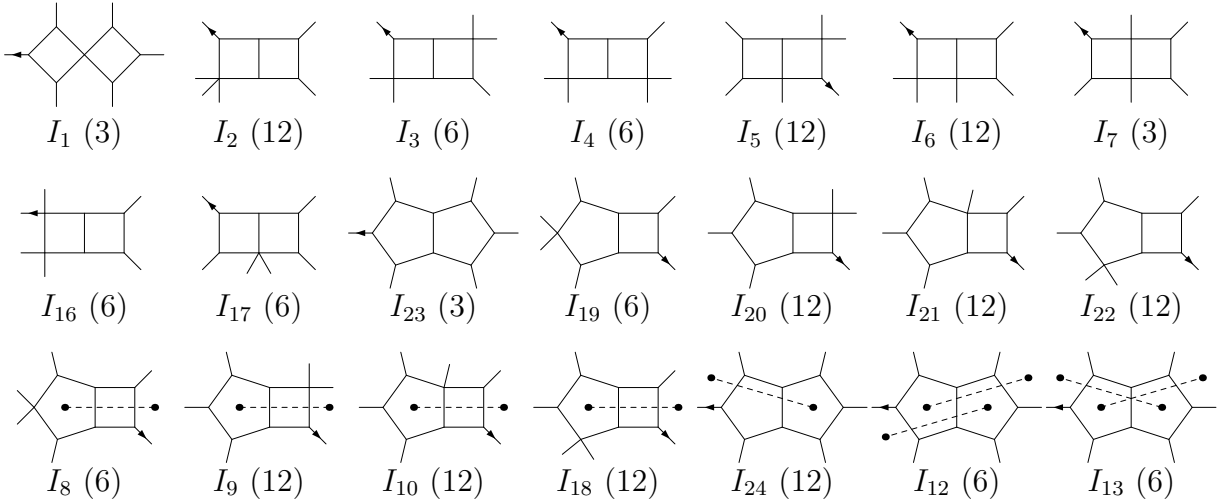}
\caption{The geometric integrals for the two-loop six-particle amplitude in
${\cal N}=4$ SYM.
In each diagram
the external momenta are labeled clockwise beginning with $k_1$ at the
position of the arrow and the number in parentheses denotes the
number of independent permutations of the diagrams.
As discussed in section IV, this is an overcomplete set: the 177
integrals here obey 18 linear relations.
}
\label{GeometricBasis}
\end{figure}

The procedure to determine the natural set of integrals in which to
represent an amplitude starts by realizing that in ${\cal N}=4$ SYM
tadpoles, bubbles,
and triangles are unneccesary (see~\cite{Bern:2006ew} for a thorough discussion).
This means that we have to start
by considering all topologies of sums over Feynman diagram with no
triangles or bubbles.
At one-loop, only sums of Feynman diagrams with the topology of
a box are needed.
For six particles
at two-loops, we find five topologies with eight propagators,
shown in~\fig{EightPropagatorTopologies}, and eight
topologies with seven propagators, shown in~\fig{SevenPropagatorTopologies}.

After each topology is identified,
a first approximation for reproducing all of the leading singularities
is to use just the scalar integrals with
all of the appropriate topologies.
In general, it turns out that such integrals are not
enough to reproduce the singularities of Feynman diagrams.
This will manifest itself in the failure of the linear
equations~(\ref{eq:method})
to have any solutions, indicating that the set of integrals must be
enlarged.

The next step
is to introduce scalar integrals with additional propagators.
At one-loop this step gives rise to pentagons in addition to boxes,
which turns out to be sufficient for any $n$.  At
two loops we add the scalar pentagon-pentagon
integrals shown in~\fig{GeometricBasis}.
At this stage
some of the equations are solved ({\it i.e.} some of the leading
singularities are correctly reproduced), while others are not.

It is then necessary to supplement
additional integrals which must have non-zero
residue on the missing singularities and zero residue on the ones which
already work, in order to avoid spoiling them.
The way to ensure that one has zero residue on a given
pole is to include a zero in the numerator of
the integrand which cancels the corresponding pole.
In this form integrals with scalar numerators appear. Note that only
numerators which cancel poles appear naturally.

The process of expanding the set of integrals
by including additional numerator
factors ends when one is able to solve all of the equations~(\ref{eq:method}).
We call the integrals that are naturally constructed in this manner
geometric integrals.
The set of geometric
integrals might be overcomplete if the equations do not determine
a unique solution.
For the six-particle MHV amplitude at two loops this process
leads to the 177 geometric integrals
shown in~\fig{GeometricBasis}.
For NMHV amplitudes it is possible that additional integrals,
such as a pentagon-boxes with two numerators, might be required.

\subsection{The Inhomogeneous Terms for the MHV Amplitude}

As mentioned above, it is obvious from~(\ref{eq:method}) that
the homogeneous part of the system of linear equations is helicity
independent, so the above discussion of the contours and
geometric integrals applies to both the MHV and NMHV configurations.
The inhomogeneous part on the right-hand side of~(\ref{eq:method})
is easily obtained for any contour $\Gamma$ by computing the product
of tree amplitudes sitting at the `blobs' in the
corresponding
topology in~\fig{EightPropagatorTopologies} or~\fig{SevenPropagatorTopologies}.
This product is evaluated at the value
$(p^{(i)}, q^{(j)})$ of the loop momenta where the contour $\Gamma$
localizes the integral.

For MHV configurations there is an enormous simplification since it
turns out that this product of tree amplitudes always comes out to be
$0$ or $1$ times the corresponding tree-level amplitude, as
can easily be shown by using the technique
introduced in~\cite{Cachazo:2008dx}
where sums over the full ${\cal N}=4$ supermultiplet in
the internal lines, which complicate the
computation~\cite{Buchbinder:2005wp,Bern:2007ct},
are automatically done by using simple
identities.

For the contours associated with the 13 topologies shown
in~\figs{EightPropagatorTopologies}{SevenPropagatorTopologies}
we find that the right-hand side of~(\ref{eq:method})
for an MHV configuration are
\begin{eqnarray}
\label{eq:RHS}
(A): &&\qquad
\delta_{\langle p,1\rangle} \delta_{\langle q,4\rangle}
A^{\rm tree}_{6,{\rm MHV}}\,, \cr
(B): &&\qquad \delta_{\langle p,6\rangle} \delta_{\langle q,1\rangle}
A^{\rm tree}_{6,{\rm MHV}}\,, \cr
(C): &&\qquad
0\,,\cr
(D): &&\qquad
\delta_{\langle p,6\rangle} A^{\rm tree}_{6,{\rm MHV}}\,, \cr
(E): &&\qquad
\delta_{\langle p,6\rangle} \delta_{\langle q,2\rangle}
A^{\rm tree}_{6,{\rm MHV}}\,, \cr
(F): &&\qquad \delta_{\langle p,6\rangle} A^{\rm tree}_{6,{\rm MHV}}\,, \cr
(G): &&\qquad 0\,,\cr
(H): &&\qquad 0\,,\cr
(I): &&\qquad \delta_{\langle p,6\rangle} \delta_{\langle p,q\rangle}
A^{\rm tree}_{6,{\rm MHV}}\,, \cr
(J): &&\qquad 0\,,\cr
(K): &&\qquad \delta_{[p,6]} \delta_{[q,1]}A^{\rm tree}_{6,{\rm MHV}}\,, \cr
(L): &&\qquad 0\,,\cr
(M): &&\qquad 0\,.
\end{eqnarray}
The Kronecker deltas such as $\delta_{\langle p,1\rangle}$
for topology $(A)$ arise
because, in that example,
the solution for $p$ is either of the form $\lambda_p \propto
\lambda_1$ or $\widetilde{\lambda}_p
\propto \widetilde{\lambda}_1$, and the product of
tree amplitudes vanishes on latter solution.

\section{The MHV Equations and Their Solutions}

We now assemble all of the ingredients prepared in section II for
the MHV amplitude.
By evaluating~(\ref{eq:method})
on all of the contours associated with the topologies
shown in~\figs{EightPropagatorTopologies}{SevenPropagatorTopologies}
(together with all of their cyclic and mirror-image permutations),
and using~(\ref{eq:RHS}) on the right-hand side,
we find a system of 396 linear equations for the 177 coefficients
of the geometric integrals in~\fig{GeometricBasis}.

Generically, 396 linear equations in 177 variables have no solution,
but in this case we find that the equations are in fact underdetermined:
they only fix 159 linear combinations of the 177 coefficients,
leaving 18 free parameters.
In other words, we find that there are 18 linear combinations of the
geometric
integrals in~\fig{GeometricBasis} which have vanishing leading
singularity on all of the contours described
by~\figs{EightPropagatorTopologies}{SevenPropagatorTopologies}.
One logically possible conclusion from such a result might have been that
the leading singularity method is not enough
to uniquely determine
the two-loop six-particle MHV amplitude, which would have been
disappointing.

Fortunately, as mentioned in the introduction, it turns out
that integral reduction identities
imply that the set of geometric
integrals is overcomplete.  In other words, there are linear combinations
of the 177 geometric integrals which not only have vanishing
leading singularity but actually vanish identically.
In section IV we analyze these reduction identities and show
that there are 18 linear combinations of the geometric integrals
which vanish, precisely accounting for the abovementioned ambiguity
in solving the leading singularity equations.
The conclusion is therefore that the two-loop six-particle MHV amplitude
is in fact uniquely determined by knowledge of its leading singularities.

\subsection{Presentation of the Equations}

In order to demonstrate that the leading singularity method
is not just black magic, we present here explicit expressions
for the equations which determine all coefficients $A^{(2)}_{6,{\rm MHV}}$,
including both the parity-even and
odd terms.
As just discussed, 18 coefficients out of 177 are actually redundant.
In order to somewhat simplify the presentation of the equations we
will choose a convenient `gauge' which uniquely fixes all of
the ambiguity.
This amounts to choosing a basis of geometric integrals.

Several such choices are possible;
the choice we make here is to spend the 18 gauge parameters
by setting to zero
the 6 coefficients $c_{12}$ and the 12
coefficients $c_{24}$.
Once this is done all remaining coefficients are uniquely determined.
A nice advantage of this choice is that several other coefficients
turn out to also vanish identically: the 12 coefficients
$c_{22}$, the 12 coefficients $c_{18}$, the 6 coefficients
$c_{16}$ and the 3 coefficients $c_{23}$
are all zero.

Ultimately then there are only 126 nonzero coefficients, associated with
just 15 out of the 21 integrals shown in~\fig{GeometricBasis}.
We label the $j$-th permutation of coefficient $c_i$ as
$c_i^{(j)}$.
Permutation $j$ maps
the labeling $(1,2,3,4,5,6)$ of the external momenta to:
\begin{eqnarray}
&& 1: (1,2,3,4,5,6)\,, \quad
   ~ 2: (2,3,4,5,6,1)\,,  \quad
    ~  3: (3,4,5,6,1,2)\,,  \quad
  ~    ~ 4: (   4,5,6,1,2,3)\,, \nonumber \\
      && 5: (5,6,1,2,3,4)\,,  \quad
         ~ 6: (6,1,2,3,4,5)\,,  \quad
        ~    7: (6,5,4,3,2,1)\,,  \quad
~~            8: (5,4,3,2,1,6)\,,  \nonumber \\
               && 9: (4,3,2,1,6,5)\,,   \quad
                  10: (3,2,1,6,5,4)\,,  \quad
                     11: (2,1,6,5,4,3)\,,  \quad
                       12: (1,6,5,4,3,2)\,.  \quad
\end{eqnarray}

Since all of the coefficients for the
MHV amplitude are proportional to the tree amplitude, we can go ahead
and divide the right-hand side of all equations by
$A^{\rm tree}_{6,{\rm MHV}}$.
Equivalently we can say that solving the equations below yields
the integral coefficients for the normalized
amplitude $A^{(2)}_{6,{\rm MHV}}/A^{\rm tree}_{6,{\rm MHV}}$.
A final comment is that we move all of the Jacobian factors (see
for example eq.~(\ref{eq:jacobian})) to the right-hand side of the
equations.

Finally we are ready to
present the equations obtained by considering the contours
associated with the topologies
in~\figs{EightPropagatorTopologies}{SevenPropagatorTopologies}.
In each equation $p$ and $q$ are understood to be evaluated at
the locations $(p^{(i)}, q^{(j)})$ of all the leading singularities.

\noindent
Topology A:
\begin{equation}
c_1^{(2)} +
\frac{
(p-k_{234})^2 (q-k_{34})^2 c_{13}^{(2)} +
(p + k_{61})^2 (q - k_{234})^2 c_{13}^{(5)}}
{(p+q-k_{234})^2}
= 4 s_{12} s_{23} s_{45} s_{56}
\delta_{\langle p,1\rangle} \delta_{\langle q,4\rangle}
\end{equation}

\noindent
Topology D:
\begin{equation}
\label{eq:TopD}
c_{19}^{(2)} + (p-k_1)^2 c_8^{(6)}  =
4 s_{12} (p - k_1)^2
(s_{123} s_{345} - s_{12} s_{45}) \delta_{\langle p,6\rangle}
\end{equation}

\noindent
Topology E:
\begin{eqnarray}
&&
c_{21}^{(3)} + (p - k_{12})^2 c_{10}^{(3)} +
\frac{
(p - k_{12})^2 (q + k_{234})^2 c_{13}^{(2)}
+ (p - k_1)^2 (q - k_{61})^2 c_{13}^{(5)}}
{(q - k_1)^2}
\cr
&&
\qquad\qquad
= 4 s_{23} s_{45} s_{56} (p - k_{12})^2
\delta_{\langle p,6\rangle} \delta_{\langle q,2\rangle}
\end{eqnarray}

\noindent
Topology F:
\begin{equation}
\label{eq:TopF}
c_2^{(6)} + \frac{c_{19}^{(2)}}{(p + k_{456})^2} = 4
s_{12}^2 s_{61} \delta_{\langle p,6\rangle}
\end{equation}

\noindent
Topology H:
\begin{eqnarray}
&&
c_4^{(6)}
+ \frac{
(p - k_{12})^2 (q - k_{56})^2 c_{13}^{(2)}
+ (p - k_1)^2 (q - k_6)^2 c_{13}^{(5)}}
{(p + k_{56})^2 (q + k_{12})^2}
\cr
&&
+ \frac{c_{20}^{(6)} + (q - k_6)^{2} c_9^{(6)}}{(q + k_{12})^2}
+ \frac{c_{20}^{(12)} + (p - k_1)^2 c_9^{(12)}}{(p + k_{56})^2}
=0
\end{eqnarray}

\noindent
Topology I:
\begin{eqnarray}
&&
c_3^{(6)} +
 \frac{
(p - k_{12})^2 (q - k_{56})^2  c_{13}^{(2)}
+ (p - k_1)^2 (q - k_6)^2 c_{13}^{(5)}}
{(p + k_{56})^2 (q + k_1)^2}
\cr
&&
+\frac{c_{20}^{(6)} + (q - k_6)^2 c_9^{(6)}}
{(q + k_1)^2}
+ \frac{c_{20}^{(3)} + (p - k_{12})^2 c_9^{(3)}}
{(p + k_{56})^2}
= 4 s_{123} (s_{123} s_{345} - s_{12} s_{45})
\delta_{\langle p,6\rangle} \delta_{\langle p,q\rangle}
\end{eqnarray}

\noindent
Topology J:
\begin{eqnarray}
\label{eq:topRMHV}
&&
c_7^{(5)}
+\frac{c_{21}^{(12)} + (p - k_{61})^2 c_{10}^{(12)}}{(p - k_6)^2}
+\frac{c_{21}^{(5)} + (q - k_{56})^2 c_{10}^{(5)}}{(q - k_6)^2}
\cr
&&
+\frac{
(p + k_{345})^2 (q - k_{56})^2 c_{13}^{(2)}
+ (p - k_{61})^2 (q - k_6)^2 c_{13}^{(5)}}
{(p - k_6)^2 (q + k_{123})^2}
\cr
&&
+\frac{c_{21}^{(2)} + (p - k_{61})^2 c_{10}^{(2)}}{(p + k_{345})^2}
+\frac{c_{21}^{(9)} + (q - k_{56})^2 c_{10}^{(9)}}{(q + k_{123})^2}
\cr
&&
+\frac{
(p - k_{61})^2 (q - k_{456})^2 c_{13}^{(1)}
+ (p - k_6)^2 (q - k_{56})^2 c_{13}^{(4)}}
{(p + k_{345})^2 (q - k_6)^2}
= 0
\end{eqnarray}

\noindent
Topology K:
\begin{eqnarray}
&& c_{17}^{(6)} +
\frac{
+ (p - k_{12})^2 (q - k_{56})^2 c_{13}^{(2)}
+(p-k_1)^2 (q-k_6)^2 c_{13}^{(5)}}
{(p+k_{456})^2(q+k_{123})^2}
\cr
&& + \frac{c_{21}^{(6)} + (q-k_6)^2 c_{10}^{(6)}}{(q+k_{123})^2}
+ \frac{c_{21}^{(12)} + (p-k_1)^2 c_{10}^{(12)}}{(p+k_{456})^2}
= 4 s_{12} s_{56} (q - k_6)^2 \delta_{\langle p,q\rangle}
\end{eqnarray}

\noindent
Topology L:
\begin{eqnarray}
&&c_5^{(3)} +
\frac{
(p - k_{12})^2 (q - k_{56})^2 c_{13}^{(2)}
+ (p - k_1)^2(q -k_6)^2 c_{13}^{(5)}}
{(p + k_{456})^2 (q + k_1)^2}
\cr
&&
+ \frac{c_{21}^{(6)} + (q - k_6)^2 c_{10}^{(6)}}{(q + k_1)^2}
+ \frac{c_{20}^{(3)} + (p - k_{12})^2 c_9^{(3)}}{(p + k_{456})^2}
= 0
\end{eqnarray}

\noindent
Topology M:
\begin{eqnarray}
&&c_6^{(6)}
+ \frac{
(p - k_{12})^2 (q - k_{56})^2 c_{13}^{(2)}+
(p-k_1)^2 (q-k_6)^2 c_{13}^{(5)}}
{(p+k_{56})^2(q+k_{123})^2}
\cr
&&
+ \frac{c_{21}^{(12)} + (p-k_1)^2 c_{10}^{(12)}}{(p+k_{56})^2}
+ \frac{c_{20}^{(6)} + (q - k_6)^2 c_9^{(6)}}{(q+k_{123})^2}
+ \frac{c_{19}^{(2)} + (p - k_1)^2 c_8^{(2)}}{(p - k_{12})^2}
=0
\end{eqnarray}

The equations for topologies $(B)$, $(C)$ and $(G)$ turn out to be
redundant for the MHV amplitude with the choice of basis
described above.

\subsection{Analytic Results for a $2 \times 2$ Block}

The structure of the equations is sufficiently complicated that it
is not clear whether it is possible to find useful analytic solutions,
so in practice we resort to solving them numerically.
However the equations from topologies ${(D)}$ and ${(F)}$
are exceptionally simple and only involve the
coefficients $c_2$, $c_8$ and $c_{19}$, so they can easily
be solved analytically as we now demonstrate.

\subsubsection{Topology $(F)$}

The four contour integrals for topology $(F)$ localize the integral
at the four points $(p^{(i)}, q^{(j)})_{i,j=1,2}$
given by
\begin{eqnarray}
q^{(1)}=\left(-\lambda_1+\frac{\langle 1,6 \rangle}{\langle 2,6 \rangle}
\lambda_2\right) \widetilde{\lambda}_1\,, && \qquad
p^{(1)}=
\frac{\langle 2,1 \rangle}{\langle 2,6 \rangle} \lambda_6
\widetilde{\lambda}_1\,
\cr
q^{(2)}=\lambda_1
\left(-\widetilde{\lambda}_1+\frac{[1,6]}{[2,6]}
\widetilde{\lambda}_2\right)\,, && \qquad
p^{(2)}=
\frac{[2,1]}{[2,6]} \lambda_1 \tilde{\lambda}_6\,.
\end{eqnarray}
Equation~(\ref{eq:TopF}) is then
\begin{equation}
\label{eq:linF}
c_2^{(6)}+\frac{c_{19}^{(2)}}{(p^{(1)}+k_{456})^2}=4 s_{12}^2 s_{61}\,, \qquad
c_2^{(6)}+\frac{c_{19}^{(2)}}{(p^{(2)}+k_{456})^2}=0\,.
\end{equation}
Note that even though there are four different contours we only obtain
two independent equations since~(\ref{eq:TopF}) is independent of $q$.
This is a generic feature whenever a contour is such that it chops off
a massless box~\cite{Cachazo:2008vp}.

Solving~(\ref{eq:linF}) yields the coefficients
\begin{equation}
\label{eq:sol19}
c_2^{(6)} = {4 s_{12}^2 s_{61}  \over 1-a}, \qquad
c_{19}^{(2)} = {4 s_{12}^2 s_{61} \over 1- a} (p^{(1)} + k_{456})^2\,
\end{equation}
where
\begin{equation}
a=\frac{(p^{(1)}+k_{456})^2}{(p^{(2)}+k_{456})^2}\,.
\end{equation}
It is frequently useful to separate coefficients into their parity-even
and parity-odd parts.  Since parity exchanges $\langle i,j\rangle
\leftrightarrow [i,j]$, it evidently takes $a \to 1/a$.
If we denote the parity conjugate of a coefficient $c$ by $\bar{c}$
then we see that
the even and odd parts of $c_2^{(6)}$ and $c_{19}^{(2)}$ are simply
\begin{eqnarray}
{1 \over 2}(c_2^{(6)}+\bar{c}_2^{(6)}) =2 s_{16} s_{12}^2\,,&& \qquad
{1 \over 2}
(c_2^{(6)}-\bar{c}_2^{(6)})=2
s_{16} s_{12}^2 \left(\frac{1+a}{1-a} \right)\,,\cr
{1 \over 2}(
c_{19}^{(2)}+\bar{c}_{19}^{(2)}) =0\,, && \qquad
{1 \over 2} (c_{19}^{(2)}-\bar{c}_{19}^{(2)}) =
4 s_{12}^2 s_{61}\frac{(p^{(1)}+k_{456})^2}{1-a}\,.
\end{eqnarray}
The parity-even parts of these coefficients
agree precisely with those obtained in~\cite{Bern:2008ap} via
the unitarity based method.  Here we see that these coefficients
can be obtained simply by solving two equations in two variables, and
moreover the parity-odd parts automatically come along for free.

\subsubsection{Topology ${(D)}$}

For topology $(D)$ the contour integrals localize the integral at the
points
\begin{eqnarray}
p^{(1)}=
(\alpha \lambda_6+\beta \lambda_3)
\widetilde{\lambda}_6\,  && \qquad
q^{(1)}= \frac{[1,2]}{[2,6]} \lambda_1 \widetilde{\lambda}_6\,,
\cr
p^{(2)}=\lambda_6 ({\alpha} \widetilde{\lambda}_6+\gamma
\widetilde{\lambda}_3)\, && \qquad
q^{(2)}=
\frac{\langle 1,2 \rangle}{\langle 2,6\rangle}
\lambda_6 \widetilde{\lambda}_1\,,
\end{eqnarray}
where
\begin{equation}
\alpha=\frac{s_{13}+s_{23}}{s_{36}},\quad
\beta=\frac{-s_{12}+\alpha (s_{12}+s_{26})}{\langle 3|1+2|6]},\quad
\gamma=\frac{-s_{12}+\alpha (s_{12}+s_{26})}{[3|1+2|6\rangle}\,.
\end{equation}
Equation~(\ref{eq:TopD}) then gives
\begin{equation}
\label{eq:linD}
c_{19}^{(2)}+\frac{c_8^{(6)}}{(p^{(1)}-k_1)^2}=
4 s_{12} (s_{345} s_{456}-s_{12} s_{45})\, \qquad
c_{19}^{(2)}+\frac{c_8^{(6)}}{(p^{(2)}-k_1)^2}=0\,.
\end{equation}
We can eliminate $c_{19}^{(2)}$ to solve for
$c_8^{(6)}$, finding
\begin{eqnarray}
{1 \over 2} (c_{8}^{(6)}  + \bar{c}_{8}^{(6)}) &=&
2 s_{12} ( s_{123} s_{345} - s_{12} s_{45})\,\cr
{1 \over 2} (c_{8}^{(6)}  - \bar{c}_{8}^{(6)}) &=&
2 s_{12} (s_{123} s_{345} -s_{12} s_{45}) \left(\frac{c+1}{c-1} \right)\,,
\end{eqnarray}
where
\begin{equation}
c=\frac{(p^{(1)}-k_1)^2}{(p^{(2)}-k_1)^2}\,.
\end{equation}
Again the even part of $c_8$ agrees precisely with the unitary based
calculation of~\cite{Bern:2008ap}.
Of course the equation~(\ref{eq:linD}) provides a consistency
condition on the coefficient $c_{19}^{(2)}$ that we already obtained
in~(\ref{eq:sol19}).

\subsection{The Parity-Even Part}

\begin{figure}
\includegraphics{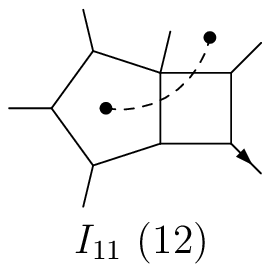}
\caption{An extra 12 integrals we add to the set of geometric
integrals in order to
facilitate comparison with~\cite{Bern:2008ap}.  These integrals are
dual conformally invariant but not geometric.  Nevertheless they
can be expressed as linear combinations of the geometric integrals
in~\fig{GeometricBasis} using first class identities (see section IV).}
\label{strangeintegral}
\end{figure}

In the previous subsection we solved for the coefficients
$c_2$, $c_8$ and $c_{19}$ analytically and demonstrated
that their parity-even parts agree with the results of~\cite{Bern:2008ap}.
In order to check the validity of the leading singularity method
it is important for us to compare the parity-even parts of all
remaining coefficients as well.  The first obstacle is the
fact that~\cite{Bern:2008ap} used a basis containing
the integral shown in~\fig{strangeintegral}.  According to our
criteria we do not consider this a geometric integral since the
propagator does not serve to cancel any pole.
(The motivation for using $I_{11}$ in~\cite{Bern:2008ap} was
that the integral is manifestly dual conformally
invariant~\cite{Drummond:2006rz,DrummondVanishing}.)

We show in the next section that reduction identities
can be used to express $I_{11}$ as a linear combination of
geometric integrals, so secretly~\fig{GeometricBasis}
already contains $I_{11}$.  However in order to facilicate comparison
of our results with those of~\cite{Bern:2008ap} it is convenient
to explicitly add $I_{11}$ to the set of integrals.  Then we have a set
of $177 + 12 = 189$ integrals which is overcomplete by
$18 + 12 = 30$ elements.
Encouragingly, we find that it is possible to find a solution
of the equations in which the non-dual conformally invariant
integrals $I_{19}$--$I_{24}$ all have coefficients whose parity-even
part vanishes.
This is a necessary condition for agreement with~\cite{Bern:2008ap}
since the parity-even part of the amplitude was expressed there
in terms of dual conformally invariant integrals.  Moreover we find
that after choosing this `gauge' there is no further ambiguity
in the basis; the linear equations furnish a unique solution.

As indicated above, the equations are sufficiently complicated that
we found it necessary to solve them numerically.
Let us note however that by `numerically' we always mean that we
choose all of the spinors $\lambda_i, \widetilde{\lambda}_i$
to be random rational numbers (subject to momentum conservation, of course).
Then the coefficients obtained by solving the equations always come
out to be rational numbers, so they can be compared to the results
of~\cite{Bern:2008ap} with absolute precision.
By repeated successful comparison for many different random values of
the spinors, we are able to report complete agreement with the
parity-even parts of the coefficients obtained from the leading
singularity method with those obtained in~\cite{Bern:2008ap},
recorded here in the $(1)$ permutation:
\begin{eqnarray}
c_1 &=&
s_{61} s_{34} s_{123} s_{345} + s_{12} s_{45} s_{234} s_{345} +
s_{345}^2 (s_{23} s_{56} - s_{123} s_{234})\,,\cr
c_2 &=&
2 s_{12} s_{23}^2\,,\cr
c_3 &=&
s_{234} (s_{123} s_{234} - s_{23} s_{56})\,,\cr
c_4 &=&
s_{12} s_{234}^2\,,\cr
c_5 &=&
s_{34} (s_{123} s_{234} - 2 s_{23} s_{56})\,,\cr
c_6 &=&
- s_{12} s_{23} s_{234}\,,\cr
c_7 &=&
2 s_{123} s_{234} s_{345} - 4 s_{61} s_{34} s_{123} - s_{12} s_{45}
s_{234} - s_{23} s_{56} s_{345}\,,\cr
c_8 &=&
2 s_{61} (s_{234} s_{345} - s_{61} s_{34})\,,\cr
c_9 &=&
s_{23} s_{34} s_{234}\,,\cr
c_{10} &=&
s_{23} (2 s_{61} s_{34} - s_{234} s_{345})\,,\cr
c_{11} &=&
s_{12} s_{23} s_{234}\,,\cr
c_{12} &=&
s_{345} (s_{234} s_{345} - s_{61} s_{34})\,,\cr
c_{13} &=&
- s_{345}^2 s_{56}\,.
\end{eqnarray}
However,
as mentioned in the introduction we note that the leading singularity
method
is completely blind to the integrals $I_{14}$ and $I_{15}$
in~\cite{Bern:2008ap} since they have integrands that vanish in $D=4$.

\section{Reductions}

\subsection{Methods of Reduction}

Repeatedly throughout the paper we have mentioned and used the new feature
that
happens for six or more particles; loop integrals often
satisfy linear
relations which can be used to write one in terms of
others (see for example~\cite{vanNeerven:1983vr},
as well as~\cite{Bern:1993kr} for dimensionally regulated versions).
Interestingly, there are relations even among what we call
geometric integrals. Also important for us will be relations among
integrals that appear in the manifestly dual conformal invariant
expression of~\cite{Bern:2008ap} and our geometric integrals.

In this
section we discuss in detail how these
relations arise since it is a crucial step in
completing the proof that the leading singularity method does determine the
amplitude uniquely. Recall that out of the $177$
coefficients the leading singularity
fixes $159$ thus leaving $18$ free parameters. We will now account for
these as a
consequence of relations or what we call reduction identities. In other
words, the set of
geometric integrals which naturally appears in the process of matching
leading singularities
is overcomplete.

We distinguish between two different kind of identities; the ones
that are valid on any contour of integration and the ones that are
valid only on $T^8$ contours where loop momenta can be taken to be
in four-dimensions. We call these first and second class identities,
respectively.

\subsubsection{First Class}

Identities of the first class are those which are valid on any
contour of integration, in particular, on the real contours where
integrals must be regulated, {\it e.g.} using dimensional
regularization.

First consider for example the pentagon-box integral
$$
\includegraphics{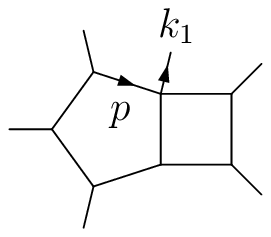}
$$
\vskip -.3cm
\noindent
with numerator factor $(p - k_1)^2$,
which is a permutation of the one shown in~\fig{strangeintegral}.
Clearly, this integral is not geometric
since the numerator is not a zero which cancels a
pole of the integral. However, this integral is dual conformal
invariant and it appears in the representation of the even part of
the amplitude obtained in~\cite{Bern:2008ap}. The goal is to write
this integral as a linear combination of the geometric
integrals in~\fig{GeometricBasis}.

Let us write the numerator as $p^2-2k_1\cdot p$. The first term
cancels a propagator and gives rise to a double-box integral of
type $I_{17}$ in~\fig{GeometricBasis}.
The second term can be decomposed by
using that external momenta are kept in four dimensions. This means
that only the four-dimensional component, $p_{[4]}$, of $p$ contributes,
{\it i.e.} $k_1\cdot p = k_1\cdot p_{[4]}$. Given any four dimensional
vector, one can write it as a linear combination of four axial
vectors
\begin{equation}
\vartheta_i^\mu =
\epsilon^{\mu\nu\rho\sigma}\ell^{i+1}_\nu\ell^{i+2}_\rho\ell^{i+3}_\sigma
\end{equation}
where $\ell^i$, with $i=1,\ldots,4$, form a basis of four-vectors.
In the case of interest, we choose to write $p_{[4]}$ as
\begin{equation}
p_{[4]}^\mu =
\frac{1}{\epsilon^{\mu\nu\rho\sigma}
\ell_\mu^1\ell^{2}_\nu\ell^{3}_\rho\ell^{4}_\sigma}
\sum_{i=1}^4 (p_{[4]}\cdot \ell_i)\vartheta_i^\mu \label{eq:verma}
\end{equation}
with the choice
\begin{equation}
\ell_1 =k_6\,, \quad \ell_2=k_5+k_6\,, \quad \ell_3 =k_4+k_5+k_6\,, \quad
\ell_4 = -k_1-k_2\,.
\end{equation}
This a standard construction in the scattering amplitude
literature~\cite{vanNeerven:1983vr,Bern:1993kr}.

Since all of the $\ell_i$ are written in terms of external particle
momenta, they are completely
four-dimensional so
we are free to replace $p_{[4]}$ by $p$ in the coefficients
$(p_{[4]} \cdot \ell_i)$.
Writing each coefficient as $2p\cdot\ell_i =
(p+\ell_i)^2-p^2-\ell_i^2$, one finds that the term
$k_1\cdot p$ in the original integral can be decomposed in terms of
numerators which give rise to geometric integrals. Perhaps the only term
which might require some
explanation is the one corresponding to $\ell_4=-k_1-k_2$ since it is one
which does not cancel
a propagator. In this case the factor $(p-k_1-k_2)^2$ becomes a numerator
which is easily seen to be a zero
that cancels a pole which removes a leading singularity
and hence gives rise to a geometric integral.

{}From this example it is clear what the necessary conditions are for
the existence of first class relations among integrals. The first condition is
that there be at least four propagators (including $1/p^2$)
involving the same loop variable and only external momenta. If the
number of such propagators is exactly four then there must be at
least two different ways of putting a numerator which only involves
the loop variable of interest and external momenta. This was the
case considered above.
If the number of
propagators is at least five, then a relation can be obtained if at
least one numerator is available. If the number of propagators is
six or more, then no numerators are needed.

In our case, with six external particles, the maximum number of
propagators in a single loop
which only depend on a single loop variable and external momenta is
five. This diagram is a hexagon-box. In fact, it is easy to show that a
hexagon-box with a
numerator can be reduced. We leave this as an exercise for the reader
since such an integral
did not have to be included in the original set of geometric integrals,
at least in the MHV case.

Using identities of the first class we will be able to show that all six
$I_{12}$ integrals can be written
in terms of $I_{13}$ integrals plus other geometric integrals.
This shows $6$ of the $18$ relations we have to account for. Also
using first class relations we will show that $6$ of the $12$ $I_{24}$
integrals can be written in terms of
the remaining six and other geometric integrals. Summarizing, after
using all first class identities we end up with
$6$ relations left to explain. These turn out to be second class identities.

\subsubsection{Second Class}

The identities discussed above are valid in any contour because the
loop momentum is always contracted with external momenta and hence
it can be treated effectively as four dimensional. We now turn to
identities which only hold on the $T^8$ contours where the loop
momentum integrals can
be taken completely in four dimensions. These identities will not
hold in dimensional regularization. In fact, their failure to hold exactly is
precisely proportional to integrals with numerators made out of the
$-2\epsilon$-dimensional
component of the loop momenta.
We obtain identities by reducing integrals of this type to find linear
combinations of other integrals which must sum to zero for
four dimensional loop momenta.

Consider first an integral with numerator, $p_{[-2\epsilon]}\cdot
q_{[-2\epsilon]} = p\cdot q - p_{[4]}\cdot q_{[4]}$.
The first term can be
written as $2p\cdot q = (p+q)^2-p^2-q^2$. In order to treat the second term
one has to find a convenient basis to expand $p_{[4]}$ and one for
$q_{[4]}$. The relevant diagram must have at least four propagators that
only depend on $p$ and at least four that only depend on $q$ and
external momenta. With six external particles there is only one
possible diagram (up to relabeling). This is the double pentagon
integral
$$
\includegraphics{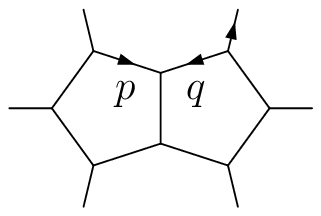}
$$
\vskip -.3cm
\noindent
Let us choose to write $p_{[4]}^\mu$ using the reference vectors
\begin{equation}
\label{eq:refa}
\ell_1 = k_6\,, \quad \ell_2 = k_5+k_6\,,\quad \ell_3 =
k_4+k_5+k_6\,,\quad \ell_4 = k_1 \label{eq:dosa}\end{equation}
while writing $q_{[4]}^\mu$ using
\begin{equation}
\label{eq:refb}
\ell_1 = k_1\,, \quad \ell_2 = k_1+k_2\,, \quad \ell_3 = k_1+k_2+k_3\,,
\quad \ell_4 =k_6 . \label{eq:huno}\end{equation}

Plugging these into~(\ref{eq:verma}) and calculating
$p_{[4]}\cdot q_{[4]}$, we find an expansion containing geometric
integrals and some further
integrals that can very easily be further decomposed into geometric ones.
These identities give rise to six relations among the remaining six
$I_{24}$ integrals and
the rest of the geometric integrals. The reader might wonder how can
one get six equations
if the starting point, which is the $p_{[-2\epsilon]}\cdot q_{[-2\epsilon]}$
pentagon-pentagon integral, has a 4-fold symmetry implying that there
are only three independent such integrals.
However, the decomposition process breaks that symmetry since one
needs to make a choice of basis vectors as in equations~(\ref{eq:refa})
and~(\ref{eq:refb}).  There are two independent choices, leading to a total
of six independent reduction identities.

One could also consider integrals with a factor of $p_{[-2\epsilon]}^2
= p^2 - p_{[4]}^2$
in the numerator.  However the only way to use a reduction
of such an integral
in the case of six particles is if $p$ is the loop momentum in a hexagon
inside the hexagon-box integral.
The hexagon-box does not appear in~\fig{GeometricBasis} since it is
never needed in order to solve the equations, so we have no
need for such reduction identities.

\subsection{Summary of All Integral Reductions}

Here we summarize all of the relevant reduction identities that can
be derived using the techniques explained above.
First we have the identity schematically represented as
$$
{\hbox{\lower 12.5pt\hbox{
\includegraphics{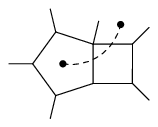}
}}} = {\hbox{\lower 12.5pt\hbox{
\includegraphics{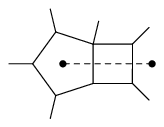}
}}} + {\hbox{\lower 12.5pt\hbox{
\includegraphics{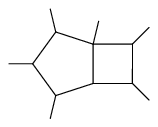}
}}} + {\rm boxes\,,}
$$
\noindent
which we use to indicate that the integral on the left can be expressed
as a linear combination of the integrals on the right.
It is straightforward to work out all of the precise coefficients,
but they are not important for our analysis.
The important point is the conclusion that the integral
shown in~\fig{strangeintegral} can be reduced to integrals
already present in~\fig{GeometricBasis}.

It is also straightforward to derive the first class relation
$$
{\hbox{\lower 12.5pt\hbox{
\includegraphics{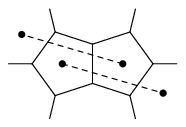}
}}}
=
{\hbox{\lower 12.5pt\hbox{
\includegraphics{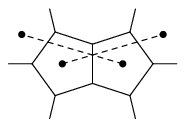}
}}}
+
{\hbox{\lower 12.5pt\hbox{
\includegraphics{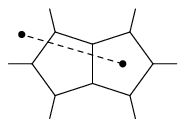}
}}}
+
{\hbox{\lower 12.5pt\hbox{
\includegraphics{int11.eps}
}}}
+
{\hbox{\lower 12.5pt\hbox{
\includegraphics{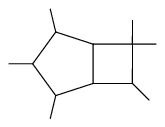}
}}}
+
{\hbox{\lower 12.5pt\hbox{
\includegraphics{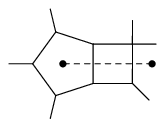}
}}}
+ {\rm boxes}\,.
$$
\noindent
This is again a schematic relation: the integrals on the right-hand side
can appear in various rotated or flipped incarnations.

The final first class relation is
$$
{\hbox{\lower 12.5pt\hbox{
\includegraphics{int24.eps}
}}}
=
{\hbox{\lower 12.5pt\hbox{
\includegraphics{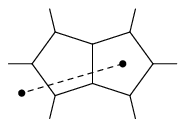}
}}}
+
{\hbox{\lower 12.5pt\hbox{
\includegraphics{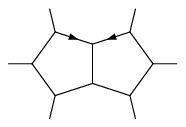}
}}}
+
{\hbox{\lower 12.5pt\hbox{
\includegraphics{int20.eps}
}}}
+
{\hbox{\lower 12.5pt\hbox{
\includegraphics{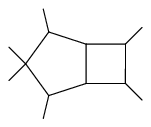}
}}}
+
{\hbox{\lower 12.5pt\hbox{
\includegraphics{int21.eps}
}}}
$$
\noindent
which implies that of the 12 apparently independent
integrals of the type $I_{24}$ shown
in~\fig{GeometricBasis}, in fact only 6 are linearly independent.

Next we summarize
the second class reduction formula, discussed in section IV.A.2, that
is obtained
by starting with a double pentagon with $p_{[-2 \epsilon]}
\cdot q_{[-2 \epsilon]}$ numerator.
As explained above this analysis leads to 6 independent identities.
Schematically these identities take the form
\begin{eqnarray}
0 &=&
{\hbox{\lower 12.5pt\hbox{
\includegraphics{int23.eps}
}}} \times p_{[-2\epsilon]}\cdot q_{[-2 \epsilon]}
\cr
&=&
{\hbox{\lower 12.5pt\hbox{
\includegraphics{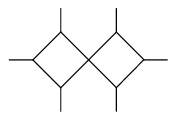}
}}}
+
{\hbox{\lower 12.5pt\hbox{
\includegraphics{int21.eps}
}}}
+
{\hbox{\lower 12.5pt\hbox{
\includegraphics{int10.eps}
}}}
+
{\hbox{\lower 12.5pt\hbox{
\includegraphics{int09.eps}
}}}
+
{\hbox{\lower 12.5pt\hbox{
\includegraphics{int20.eps}
}}}
+
{\rm boxes}\,.
\nonumber
\end{eqnarray}
\noindent
Again the integrals on the right-hand side can appear in various
different permutations.

Taking everything into account, we find that the set of
geometric integrals in~\fig{GeometricBasis} is overcomplete by
$6 + 6 + 6 = 18$ elements.  This precisely accounts for the 18 free
parameters we found in solving the linear equations for the MHV coefficients.
(This increases from 18 to $18 + 12 = 30$, as discussed in section III.C,
when the integral $I_{11}$ is thrown into the mix.)

\section{Numerical Check of the ABDK/BDS Ansatz}

One of the most interesting properties of scattering
amplitudes in ${\cal N}=4$ SYM is that the structure of infrared
divergences in higher loop amplitudes
is very simply related to those of lower loop amplitudes~\cite{KnownIR}.
In~\cite{ABDK,BDS}, it was conjectured that this simplicity
persists, at least for MHV
amplitudes, to the finite terms as well.
The precise form of the conjecture at two loops, in dimensional
regularization to $D = 4 - 2 \epsilon$, is
\begin{equation}
\label{eq:jaso}
M_{n,{\rm MHV}}^{(2)}(\epsilon)  =
\frac{1}{2}M_{n,{\rm MHV}}^{(1)}(\epsilon)^2
+ f^{(2)}(\epsilon) M_{n,{\rm MHV}}^{(1)}(2\epsilon) - {\pi^4 \over 72} +
{\cal O}(\epsilon)\,,
\end{equation}
where $M^{(L)}_{n,{\rm MHV}}
= A^{(L)}_{n,{\rm MHV}}/A^{\rm tree}_{n,{\rm MHV}}$
is the normalized $L$-loop amplitude
and
$f^{(2)}(\epsilon) = -(\zeta(2) + \zeta(3) \epsilon + \zeta(4) \epsilon^2 +
\cdots)$.  The conventions implicit in equation~(\ref{eq:jaso})
require that every loop momentum integral $p$ be normalized with the factor
\begin{equation}
-i \pi^{-D/2} e^{\epsilon \gamma} \int d^D p\,.
\end{equation}

The simple structure~(\ref{eq:jaso}) holds perfectly for $n=4$ and
$n=5$~\cite{ABDK,TwoLoopFiveA,TwoLoopFiveB}.
However it apparently fails beginning at $n=6$ particles. This was found
in~\cite{Bern:2008ap} by
computing the parity-even part of $M^{(2)}_{6,{\rm MHV}}$
numerically and finding disagreement with the right-hand side
of~(\ref{eq:jaso}).
Here we do not have anything to add to this issue
except that the parity-even piece of our full answer agrees with the
result of~\cite{Bern:2008ap}, thus providing independent confirmation.

Since the leading singularity method allows us to obtain the parity-odd
parts of all coefficients with no more effort than
the parity-even parts, we are in a position to test
the parity-odd part of~(\ref{eq:jaso}) for
$n=6$.  Note that the loop momentum integrals must necessarily be
evaluated numerically (except for $n=4$, where analytic results are
known through three loops) with current state-of-the-art technology
(in particular we use~\cite{MB,CUBA}).

Restricting to the parity-odd part of~(\ref{eq:jaso}) yields
\begin{equation}
\label{eq:oddabdk}
M^{(2)}_{6, {\rm MHV~odd}}(\epsilon) = M^{(1)}_{6, {\rm MHV~even}}(\epsilon)
M^{(1)}_{6, {\rm MHV~odd}}(\epsilon) + {\cal O}(\epsilon).
\end{equation}
The one-loop amplitude is~\cite{BDDKSelfDual},
\begin{eqnarray}
M^{(1)}_{6, {\rm MHV~even}}(\epsilon)
&=& - {1 \over 2 \epsilon^2} \sum_{i=1}^6 (-s_{i,i+1})^{-\epsilon}
+ {\cal O}(1)\,,\cr
M^{(1)}_{6, {\rm MHV~odd}}(\epsilon)
&=&
-{1 \over 4} \sum_{i=1}^6
\Big(
\langle i|i{+}1|i{+}2|i{+}3|i{+}4]
-
[i|i{+}1|i{+}2|i{+}3|i{+}4\rangle
\Big) P_{i{+}5,i{+}6}(\epsilon)\,,
\end{eqnarray}
where $P_{i{+}5,i{+}6}(\epsilon)$ is the one-loop pentagon integral
with external legs $i{+}5$ and $i{+}6$ joined
and with a factor of $p_{[-2\epsilon]}^2$ in the numerator.

We have evaluated~(\ref{eq:oddabdk}) numerically at two independent
kinematic points with randomly generated values of $\lambda_i$
and $\widetilde{\lambda}_i$ for the six external particles.
Denoting the left- and right-hand sides of~(\ref{eq:oddabdk})
by $L$ and $R$ respectively, we find
\begin{eqnarray}
L(\epsilon) &=& - \frac{4 \times 10^{-16}}{\epsilon^4}
+ {4 \times 10^{-15} \over \epsilon^3}
+ \frac{1(2) \times 10^{-11}}{\epsilon^2}
- \frac{0.430(7)}{\epsilon}
- 0.9(1) + {\cal O}(\epsilon)\,,\cr
R(\epsilon) &=&\qquad\qquad\qquad\qquad~~~~~~~~~~~~~~~~~~~~~~~~~~~~
- \frac{0.428(2)}{\epsilon} -
0.92(1) + {\cal O}(\epsilon)
\label{eq:res1}
\end{eqnarray}
at the first point and
\begin{eqnarray}
L(\epsilon) &=& { < 10^{-16} \over \epsilon^4}
- {8 \times 10^{-15} \over \epsilon^3}
+ {7(5) \times 10^{-12} \over \epsilon^2}
- {15.902(5) \over \epsilon}
- 60.46(6) + {\cal{O}}(\epsilon)\,,
\cr
R(\epsilon) &=&\qquad\qquad ~~~~~~~~~~~~~~~~~~~~~~~~~~~~~~~~~~~~
- \frac{15.915(6) }{\epsilon}
- 60.38(4) + {\cal{O}}(\epsilon)
\label{eq:res2}
\end{eqnarray}
at the second.
In these expressions the value in parentheses denotes the estimated
numerical error in the last digit as reported by~\cite{CUBA}.

We emphasize that the cancellation of the divergent terms
in~(\ref{eq:res1}) is not a check of the ABDK/BDS
conjecture, but rather a check on our application of the leading
singularity method.  This is so because it is a known fact~\cite{KnownIR},
not a conjecture, that~(\ref{eq:jaso}) must hold for the
infrared divergent terms.
Had we gotten a nonzero result, it would have signalled an error in
our calculation of the integral coefficients.
Note that the cancellation in $L(\epsilon)$ is highly nontrivial
in the sense that the result shown is obtained after summing of order
100 contributions which are typically of order~$1$.
It is the fact that we see the cancellation persisting to order
${\cal{O}}(\epsilon^0)$ that strongly suggests that the parity-odd part
of the two-loop six-particle MHV amplitude indeed satisfies the
ABDK/BDS
conjecture~(\ref{eq:oddabdk}).

However it is important to note once again that since the leading
singularity method is not sensitive to any ``$\mu$''-terms we
cannot rule out the possibility that there may be additional
such contributions to $M^{(2)}_{6,{\rm MHV~odd}}$.
Given our apparently successful check of~(\ref{eq:oddabdk}) there
are three possibilities: (1) the two-loop amplitude
does not contain any parity-odd $\mu$-terms (this is indeed
the case for $n=5$ particles~\cite{TwoLoopFiveB}), (2) the amplitude
does contain parity-odd $\mu$-terms but they contribute
only at ${\cal O}(\epsilon)$, or (3) the amplitude contains
parity-odd $\mu$-terms which spoil the ABDK/BDS relation~(\ref{eq:jaso}).

\section{Conclusion}

In this paper we employed the leading singularity method to
determine the integral coefficients of the planar two-loop six-particle
MHV amplitude in ${\cal N} = 4$ YM, the parity-even parts of which
were recently obtained in~\cite{Bern:2008ap} using the unitarity based
method.
One advantage of the leading singularity method is that the full
coefficients, including parity-odd parts, emerge from solving
the relatively simple set of linear equations displayed explicitly
in section III.A.

The leading singularity method has previously proven
succesful~\cite{Cachazo:2008vp}
at reproducing the $n=4$~\cite{Bern:1997nh} and
$n=5$~\cite{TwoLoopFiveB}
particle amplitudes
at two loops.
However,
there is currently no proof
that a general amplitude in ${\cal N} = 4$ Yang-Mills is uniquely
determined by its leading singularities only.
It is a logical possibility that finding a representation of an
amplitude in terms of simpler integrals which faithfully reproduce
all of the leading singularities is not a sufficient condition
to guarantee correctness of the representation, although clearly
it is a necessary condition.

Here we find that the $n=6$ particle amplitude is in fact completely
determined by its leading singularities, although establishing
this fact required that considerable attention be given
to the choice of basis for
the integrals and reduction identities which relate various integrals
to each other.  This was necessitated by the fact that the full set
of linear equations we found does not have a unique solution.
Fortunately we found that all of the ambiguities could be accounted
for by taking into account reduction identities.

One nice feature of the leading singularity method is that the procedure
naturally provides a set of `geometric' integrals
for any amplitude under consideration.
The set of geometric integrals does not coincide with the manifestly
dual conformally invariant~\cite{Drummond:2006rz,DrummondVanishing}
basis used to express the parity-even part of the $n=6$ amplitude
in~\cite{Bern:2008ap}.
We expect
this to be true in general.
This is not surprising given the fact
that by using the leading singularity technique both even and odd parts
of the amplitude are computed simultaneously, whereas
already for five particles the odd part of the
amplitude~\cite{TwoLoopFiveB}
is not expressible in terms of integrals with manifest dual conformal
properties.  Our results indicate that also for $n=6$ the odd
part of the amplitude cannot be expressed in terms of dual conformally
invariant integrals alone.

Another motivation for computing the MHV six-particle amplitude,
beyond its serving as a testing ground for the
leading singularity method,
is to study the so-called ABDK/BDS
conjecture for MHV amplitudes~\cite{ABDK,BDS}.
Although the conjecture was shown to be violated by the
parity-even part of the $n=6$ particle amplitude~\cite{Bern:2008ap},
following earlier doubts that had been raised
in~\cite{AMTrouble,BNST,Lipatov}, we provide numerical evidence
that the parity-odd part of the amplitude does satisfy
the ABDK/BDS relation.

Equivalently, one can say that the parity-odd
part evidently cancels out when one takes the logarithm of
the resummed amplitude (see~\cite{BDS}).  Although we
do not know of any proof that this has to be the case, the result
is consistent with the structure
seen at strong coupling~\cite{AM}, which is manifestly parity-invariant.
It is also consistent with the astounding but still mysterious
equivalence between
scattering amplitudes and lightlike Wilson loops in ${\cal{N}} = 4$ YM
that has been observed at one-loop~\cite{DrummondVanishing,
BrandhuberWilson} and at two-loops through $n=6$
particles~\cite{DHKSTwoloopBoxWilson,
ConformalWard,HexagonWilson,Drummond:2008aq}, since the
`vanilla' Wilson loop does not carry any helicity information.
It is a very interesting open problem to determine if the
amplitude/Wilson loop equivalence can be extended to other helicity
configurations by appropriately dressing the Wilson loop.

\section*{Acknowledgments}

M.~S. and A.~V. are grateful to Z.~Bern,
L.~J.~Dixon,
D.~A.~Kosower,
R.~Roiban,
C.~Tan,
C.~Vergu and
C.~Wen for many helpful comments and collaboration on related questions.
We thank D.~A.~Kosower for sharing
Mathematica files containing some results
from the unitarity based
method.
The authors are also grateful to
the Institute for Advanced Study for hospitality
during the origination of this work.
This work was supported in part by the
NSERC of Canada and MEDT of Ontario (F.~C.), the
US Department of Energy under contract
DE-FG02-91ER40688 (M.~S. (OJI) and A.~V.),
the US National Science Foundation
under grants PHY-0638520 (M.~S.) and PHY-0643150
CAREER (A.~V.).

\end{document}